\documentstyle[12pt,epsfig]{article}
\newcommand{\bm}[1]{\boldmath{\mbox{$ #1 $}}}
\newcommand{\fnd}[2]{\frac{\displaystyle #1}{\displaystyle #2}}
\begin{document}
\title{\bf
The nature of \bm{\sigma}, \bm{\kappa}, \bm{a_0(980)} and \bm{f_0(980)}}
\author{
E.\ van Beveren\footnote{Electronic address: eef@teor.fis.uc.pt} \\
{\normalsize\it Centro de F\'{\i}sica Te\'{o}rica, Departamento de F\'{\i}sica,
Universidade de Coimbra} \\ {\normalsize\it P-3004-516 Coimbra, Portugal}\\[3mm]
\and
D.\ V.\ Bugg\footnote{Electronic address: D.Bugg@rl.ac.uk} \\
{\normalsize\it Queen Mary, University of London, London E1\,4NS, UK} \\[3mm]
\and
F.\ Kleefeld\footnote{Present address: Doppler Institute for Mathematical 
Physics and Applied Mathematics \& Nuclear Physics Institute, Czech Academy of
Sciences, 250\,68 \v{R}e\v{z}, Czech Republic; electronic address:
kleefeld@cfif.ist.utl.pt}\,\, and
G.\ Rupp\footnote{Electronic address: george@ist.utl.pt}  \\
{\normalsize\it Centro de F\'{\i}sica das Interac\c{c}\~{o}es Fundamentais,
Instituto Superior T\'{e}cnico}\\ {\normalsize\it P-1049-001 Lisboa, Portugal}}
\date{\today}
\maketitle

\begin{abstract}
Masses and widths of  the four light scalar mesons $\sigma$,
$\kappa$, $a_0$(980) and $f_0$(980) may be reproduced in a model
where mesons scatter via a $q\bar q$ loop.
A transition potential is used to couple mesons to $q\bar q$ at a
radius of $\sim 0.57$ fm.
Inside this radius, there is an infinite bare spectrum of
confined $q\bar{q}$ states, for which a harmonic oscillator is chosen
here.
The coupled-channel system approximately reproduces the features of
both light and heavy meson spectroscopy.
The generation of $\sigma$, $\kappa$, $a_0(980)$ and $f_0(980)$ is a
balance between attraction due to the $q\bar q$ loop and
suppression of the amplitudes at the Adler zeros.
Phase shifts increase more rapidly as the coupling constant to the
mesons increases.
This leads to resonant widths which decrease with increasing coupling
constant --- a characteristically non-perturbative effect. \\[2mm]
\small PACS numbers: 14.40.Cs, 14.40.Ev, 13.25.-k, 13.75.Lb
\end{abstract}

Recent experiments have improved parameters of $\sigma$, $\kappa$,
$a_0(980)$ and $f_0(980)$ \cite{Ai01}--\cite{Ab05}, and there are now
extensive data on their couplings to decay channels.
Their masses do not conform to the pattern set by well-known
$q\bar q$ nonets, e.g. $\rho$, $\omega $, $K^*(890)$ and
$\phi$.
The mass of the $a_0(980)$ is $\sim 450$ MeV above the mass of
the $\sigma$, unlike the near degeneracy between $\rho$ and $\omega$.
This has led Jaffe to suggest that they are 4-quark states \cite{J77}.
However, Scadron has shown that the existence of a light scalar nonet
may be a direct consequence of dynamical chiral-symmetry breaking
\cite{S82}.

Here, we present a simple model which reproduces their essential
features.
It fits $\pi \pi$ and $K\pi$ elastic phase shifts, and also the
line-shapes of $a_0(980) $ and $f_0(980)$, i.e.,
$a_0 \to \pi \eta$, $KK$ and $\pi \eta '$, and
$f_0 \to \pi \pi$, $KK$ and $\eta \eta$.

The model  in its two versions has been described in earlier
publications \cite{BRMDRR86}\cite{BR01}.
In one formulation \cite{BRMDRR86}, an explicit harmonic oscillator
potential, with corresponding wave functions, is used for the bare
$q\bar{q}$ states; it reproduces the gross features of the spectra
of both light and heavy mesons.
In the other approach \cite{BR01}, use is made of a
so-called \em Resonance Spectrum Expansion, \em which allows, in
principle, the
use of any confinement spectrum for the bare states.
In the present paper, the latter method is applied, though again with
a harmonic oscillator.
This choice is not crucial, as e.g.\ the $f_0(1370)$ lies well above
states considered here and couples mostly to $4\pi$.
So the more conventional funnel potential is expected to give
similar results.

\begin{figure}[htb]
\begin{center}
\centerline{\hspace{0.5cm}\epsfig{file=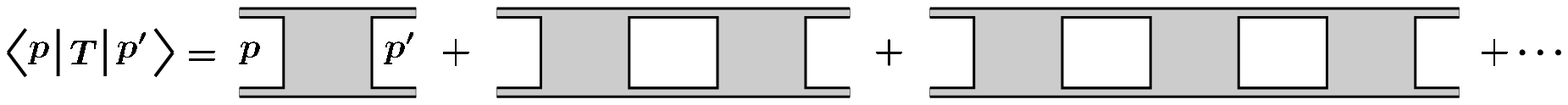,width=17cm}}
\end {center}
\caption{Schematic picture of the model.}
\label{fig1}
\end {figure}

The confining potential is joined at radius $r_0$ to plane waves for
meson pairs.
Decays of $q\bar q$ states trapped in the confining potential are
described by Fig.~\ref{fig1}.
The $\sigma$, $\kappa$, $f_0(980)$ and $a_0(980)$ appear as extra
states through their dynamical coupling to this $q\bar q$ loop.
If it were not for this coupling, they would be plane waves in the
continuum.
Coupling of mesons to the loops induces some degree of mixing
of regular $q\bar q$ states with $\sigma$ and its relatives.

A universal coupling constant $\lambda$ is used
for quarks of all flavours, after a mass scaling as a
function of the reduced quark mass ensuring approximate flavour
blindness of our equations, provided the same scaling is applied
to the radius $r_0$ \cite{BR04}.
The relative 3-meson couplings at each vertex are taken from
Tables I and II of Anisovich {\it et al.} \cite{AKPSS00}.
These are the usual $SU(3)$-symmetric couplings for OZI-allowed
decays, to be contrasted with the totally flavour-symmetric choice used in
the second paper of Ref.~\cite{BRMDRR86}. The latter convention gives rise
to equally good fits, but results in a significantly larger value of $\lambda$
for the coupled $\sigma$-$f_0$(980), as compared to the $\kappa$ and
$a_0$(980), in the present restriction to pseudoscalar-pseudoscalar (PP)
decay channels only. The formulae allow for the pseudoscalar mixing angle
$\Theta_{PS}$ in the couplings of $\eta $ and $\eta'$,
\begin {equation}
\begin{array}{r}
\eta \; = \; n\bar{n}\cos(\Theta_{0}+\Theta_{PS})-
             s\bar{s}\sin(\Theta_{0}+\Theta_{PS}) \; , \\[1mm]
\eta'\; = \; n\bar{n}\sin(\Theta_{0}+\Theta_{PS})+
             s\bar{s}\cos(\Theta_{0}+\Theta_{PS}) \; ,
\end{array}
\end {equation}
where $\Theta_{0}=\arctan\sqrt2\approx 54.7^\circ$, thus defining $\Theta_{PS}$
in the octet-singlet basis.
The intrinsic (or bare) mixing between $\sigma$ and $f_0(980)$ is likewise
expressed, this time in the $n\bar{n}$-$s\bar{s}$ basis, in terms of a scalar
mixing angle $\Theta_S$ by
\begin{equation}
\begin{array}{r}
\sigma \; = \; n\bar{n}\cos\Theta_S-s\bar{s}\sin\Theta_S \; , \\[1mm]
f_0    \; = \; n\bar{n}\sin\Theta_S+s\bar{s}\cos\Theta_S \; .
\end{array}
\end{equation}
Coupled relativistic Schr\"{o}dinger equations are solved outside
and inside the radius $r_0$, and are matched at that radius.
The resulting multichannel $T$ matrix can be written down in closed form.
In the case of the $\kappa$ and the $a_0$(980), as well as the decoupled
$\sigma$ and $f_0$(980), there is only one $q\bar{q}$ channel,
resulting in the relatively simple expression
\begin{equation}
T_{ij}(E)\; =\;
-\fnd{2r_0\lambda^2\left\{\sum_{s=0}^{\infty}\frac{g_i(s)g_j(s)}{E-E_s}\right\}
\sqrt{\mu_i\mu_jk_ik_j}\,j_{\ell_i}(k_i r_0)\,
j_{\ell_j}(k_j r_0)}{1\,+\,2ir_0\lambda^2\sum_{n=1}^{N_f}
\left\{\sum_{s=0}^{\infty}\frac{g_n^2(s)}{E-E_s}\right\}\mu_n k_n \,
j_{\ell_n}(k_n r_0)\,h^{(1)}_{\ell_n}(k_n r_0)} \; ,
\label{Tij}
\end{equation}
where $\lambda$ is the overall coupling, $r_0$ is the delta-shell radius,
$g_i(s)$ is the relative coupling of the $i$th ($i=1,N_f$) two-meson channel,
depending on the radial excitation (see Ref.~\cite{BR01}, Table 1 of 4th
paper), $E_s$ are the levels of the discrete confinement spectrum,
$\mu_i=\mu_i(E)$ and $k_i=k_i(E)$ are the relativistic
reduced mass and momentum in the $i$th channel, and $j_{\ell_i}$ and
$h^{(1)}_{\ell_i}$ are spherical Bessel and Hankel functions, respectively.
For the coupled $\sigma$ and $f_0$(980), there are two $q\bar{q}$ channels,
i.e., $n\bar{n}$ and $s\bar{s}$, giving rise to a more complicated expression.
Writing $T_{ij}(E)=N_{ij}(E)/D(E)$, we have
\begin{eqnarray}
\lefteqn{N_{ij}(E) \; =\; -2r_0\lambda^2\sqrt{\mu_i\mu_j k_i k_j}\,
j_{\ell_i}(k_i r_0)\,j_{\ell_j}(k_j r_0) \left(\rule{0cm}{1cm} 
\sum_{\alpha=1}^{2}\left\{\sum_{s=0}^{\infty}\frac{g_{\alpha i}(s)
g_{\alpha j}(s)}{E-E_{\alpha s}}\right\}\right. \; + \;} \\[2mm]
& + & \left.2ir_0\lambda^2\sum_{n=1}^{N_f}\left\{\sum_{s,s'=0}^{\infty}
\raisebox{-2mm}{$
\fnd{\left| \begin{array}{cc} g_{1i}(s) & g_{1n}(s) \\
                               g_{2i}(s')& g_{2n}(s')    \end{array}\right|
      \left| \begin{array}{cc} g_{1j}(s) & g_{1n}(s) \\
                               g_{2j}(s')& g_{2n}(s')    \end{array}\right|}
{\rule{0cm}{5mm}(E-E_{1s})(E-E_{2s'})}$} \right\} \mu_n k_n \,
j_{\ell_n}(k_n r_0)\,h^{(1)}_{\ell_n}(k_n r_0)\right) 
\nonumber\label{Nij}
\end{eqnarray}
and
\begin{eqnarray}
\lefteqn{D(E) \; =\; 1+2ir_0\lambda^2 \sum_{\alpha=1}^{2}\sum_{n=1}^{N_f}
\left\{\sum_{s=0}^{\infty}\frac{g^2_{\alpha n}(s)}{E-E_{\alpha s}}\right\}
\mu_n k_n \, j_{\ell_n}(k_n r_0)\,h^{(1)}_{\ell_n}(k_n r_0)\; + \;} \\[2mm]
&\!\!-\!\!&2r^2_0\lambda^4\!\!\sum_{\stackrel{\scriptstyle n,n'=1}{n\neq n'}}^
{N_f} \left\{\sum_{s,s'=0}^{\infty} \raisebox{-2mm}{$
\fnd{ \left| \begin{array}{cc} g_{1n}(s) & g_{1n'}(s) \\
                               g_{2n}(s')& g_{2n'}(s')    \end{array}\right|^2}
{\rule{0cm}{5mm}(E-E_{1s})(E-E_{2s'})}$} \right\}
\mu_{n }k_{n }\,j_{\ell_{n }}(k_{n } r_0)\,h^{(1)}_{\ell_{n }}(k_{n } r_0)\,
\mu_{n'}k_{n'}\,j_{\ell_{n'}}(k_{n'} r_0)\,h^{(1)}_{\ell_{n'}}(k_{n'} r_0) \; ,
\nonumber\label{D}
\end{eqnarray}
where the index $\alpha=1,2$ refers to the $n\bar{n}$ and $s\bar{s}$
confinement channels, respectively.

The formulae of Anisovich \em et al.\ \em allow for separate coupling
constants for $s\bar s$ and $n\bar n$; we find this refinement does not
give any significant improvement, so we use the same value of
$\lambda$ for both.

A key feature of the model lies in the automatic inclusion of  Adler zeros
in scattering amplitudes,  which results from using relativistic reduced
masses for two-meson channels \cite{RKB05}.
It leads to Adler zeros at $s = 0$ for $\pi \pi$ and $KK$, at
$s = m^2_\eta - m^2_\pi$ for $\eta \pi$, and at $s = m^2_K - m^2_\pi$
for $K\pi$; although these are not quite conventional values,
e.g.\ $0.5\,m^2_\pi$ for $\pi\pi$, the difference is insignificant for
present purposes.
The Adler zeros are essential to reproduce $\pi \pi$ and $K\pi$
elastic phase shifts near threshold \cite{B03}.

Experimental data are used to determine the coupling constant
$\lambda$ and transition radius $r_0$.
The optimum values turn out to be in rough agreement with those in
previous work \cite{BR01}\cite{BR04} (see below).
Phase shifts for $\pi \pi \to \pi \pi$ are taken from
the BES publication \cite{Ab04} on the $\sigma$ pole (Method 1), where
a combined fit was made to (i) BES data on $J/\Psi \to \omega \pi ^+\pi
^-$;  (ii) Cern-Munich data \cite{H73}; (iii) $K_{e4}$ data of Pislak
\em et al.\ \em \cite{P01}; (iv) the $\pi \pi$ scattering length
$a_0 = 0.220 \pm 0.005$ determined from chiral perturbation theory (ChPT) by
Colangelo, Gasser and Leutwyler \cite{CGL01}.
Phase shifts for $K\pi \to K\pi$ are taken from the combined fit
\cite{B06_1} to LASS data for $K\pi $ elastic scattering \cite{A88},
and E791 \cite{Ai06} and BES data \cite{B05}\cite{Ab06} for the
$\kappa $ pole; because of uncertainties in fitting
$K_0^*(1430)$ (discussed below), the fit is made from threshold to 1.2
GeV after subtracting phases due to $K_0^*(1430)$.
The line-shape of $a_0(980)$ is taken from the recent combined
analysis \cite{B06_2} of KLOE data for $\phi \to \gamma a_0$
\cite{Al02}, and Crystal Barrel data on $\bar pp \to \omega a_0$ and
$a_0\pi $ \cite{BASZ94}.
The line-shape of $f_0(980)$ is taken from BES data \cite{Ab05}.

The coupling constant $\lambda$ and radius $r_0$ are fitted for each
scalar resonance separately, except for the $f_0(980)$.
We also make a combined fit to $\sigma$ + $f_0(980)$, in which both
states emerge dynamically, by coupling the bare isoscalar $n\bar{n}$
and $s\bar{s}$ channels to $\pi\pi$, $KK$, $\eta\eta$, $\eta\eta'$
and $\eta'\eta'$. Note that, although there is no direct coupling of
$s\bar{s}$ to $\pi\pi$, the higher-order OZI-allowed process
$s\bar{s}\to KK\to n\bar{n}\to\pi\pi$ does produce a coupling
after all. This also produces a dynamical mixing of $n\bar{n}$ and
$s\bar{s}$, in addition to the intrinsic quark-level mixing assumed
above. Results for parameters and pole positions are shown in
Table~\ref{table1}. 
\begin{table}[htb]
\begin {center}
\begin{tabular}{cccc}
\hline
Resonance & $\lambda $ & $r_0$  & Pole Positions \\
 & (GeV$^{-3/2}$) & (GeV$^{-1}$)   & (MeV) \\\hline
$\sigma$   & 2.92 & 2.84 &  555 - i262 \\
$\kappa $  & 2.97 & 3.29 &  745 - i316 \\
$a_0(980)$ & 2.62 & 2.81 &  1021 - i47 \\
$\sigma + f_0(980)$ & 3.11& 2.71&  530 - i226 , 1007 - i38\\ \hline \\[-9mm]
\end{tabular}
\end{center}
\caption{Fitted parameters. For $a_0$ and $f_0$, second-sheet poles
are quoted.}
\label{table1}
\end{table}
Striking is the considerable effect on the $\sigma$ pole from coupling to
the $f_0$(980), showing the influence of $n\bar{n}$-$s\bar{s}$ mixing via
the kaon loop. As for the parameters, there is a reasonable agreement for
all cases.
The $\lambda$ parameter varies by $\pm 9\%$ and the radius parameter
by $\pm 10\%$. Moreover, both parameters are roughly compatible
with the values $\lambda=0.75$ GeV$^{-3/2}$ and $r_0=3.2$ GeV$^{-1}$
used in previous model calculations with only one continuum channel
\cite{BR01}\cite{BR04}, taking into account that in the present study
the \em squared \em \/relative couplings of the included
PP channels are normalised so as to add up to
1/16, whereas in the single-channel case this number was just set to 1.
Some small variations of parameters are to be expected due to the choice
of a delta shell for the transition potential, which is very sensitive to
the precise radial wave functions of the different PP channels.
We shall comment on global features first and return to detail later.

\begin{figure}[htb]
\mbox{} \\[-1cm]
\begin {center}
\centerline{\hspace{-2cm}\epsfig{file=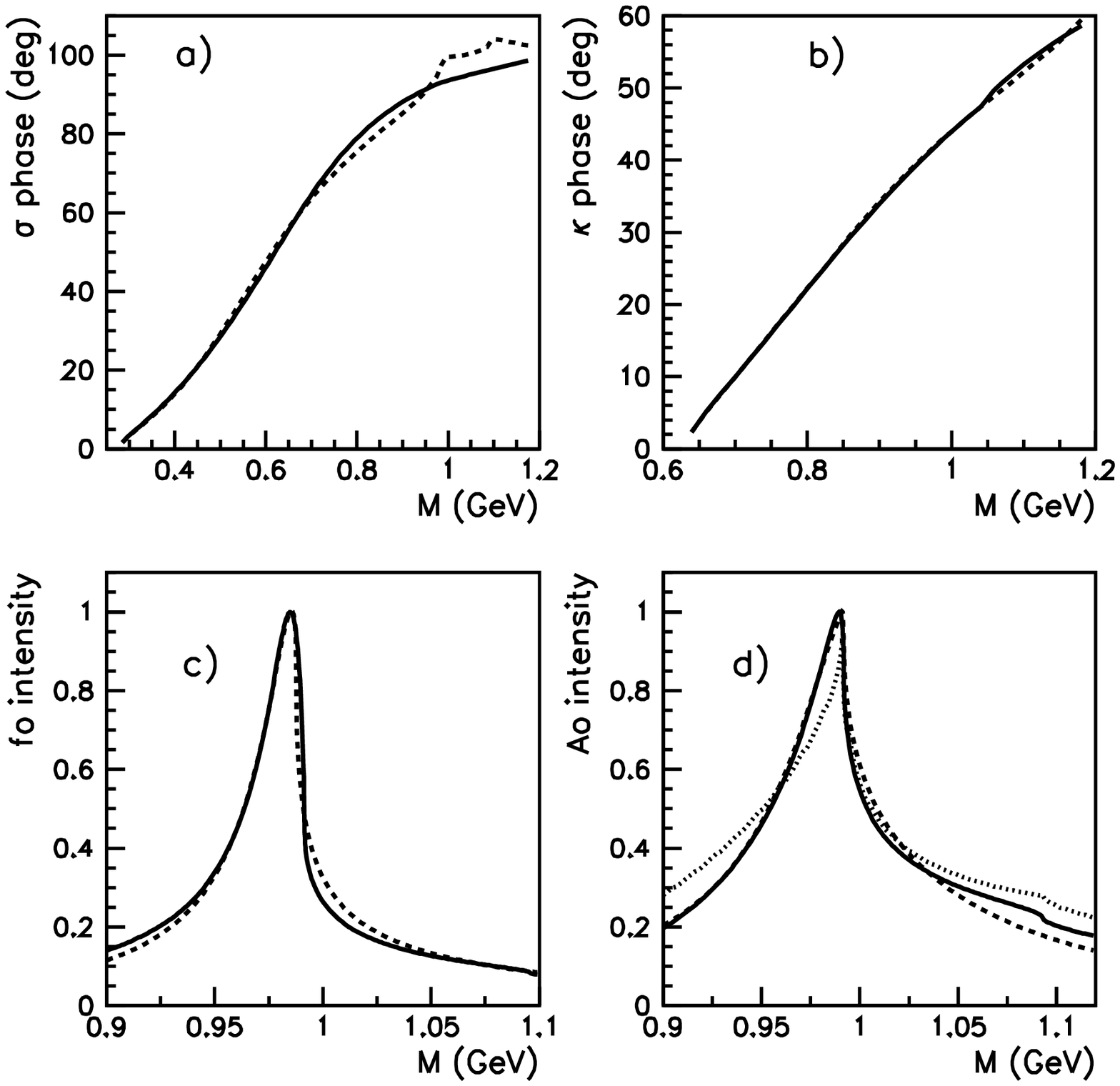,width=15cm}}
\end {center}
\mbox{} \\[-2.5cm]
\caption{Fits to (a) the $\sigma$ component of the $\pi \pi$
phase shift for elastic scattering; (b) the $\kappa$ component of
$K\pi$ phase shifts for elastic scattering;
(c) + (d) the line shapes of $f_0(980)$ and $a_0(980)$ in
$\pi \pi$ resp.\ $\pi \eta$ channels.}
\label{fig2}
\end {figure}
In Fig.~\ref{fig2}(a), the dashed curve shows our fit to
the BES fit (full curve) to $\pi \pi$ phase shifts after subtracting the
$f_0(980)$ component.
Structure in our fit near the $KK$ and $\eta \eta $ thresholds
originates from $\sigma \to KK$ and $\eta \eta$.
These thresholds were not included in the analysis of BES data,
so the comparison illustrates the magnitude of possible effects
from those thresholds.
There is also structure in our fit from 800 to 1000 MeV,
caused by quantum-mechanical mixing between $f_0(980)$ and the $\sigma$
amplitude.
Figure~\ref{fig2}(b) shows the fit (dashed) to $K\pi$ phase shifts up to
1.2 GeV from LASS (full curve) after subtracting the component due to
$K_0^*(1430)$.
Note that our model predicts a small cusp effect at the $K\eta$
threshold (also see Ref.~\cite{K05b}), but within the errors of the data
(typically $\pm 3^\circ$). Figures~\ref{fig2}(c) and \ref{fig2}(d) show fits
to the line-shapes of $f_0(980)$ and $a_0(980)$, respectively.
The original fits \cite{B06_2} to experimental data were made using Flatt\'{e}
formulae with widths of the form $g^2\rho$ for each channel, where
$\rho$ is phase space. Our model is then fitted to these parametrisations,
giving slightly more complicated line-shapes.
Also notice the cusp at the $\eta'\pi$ threshold in Fig.~\ref{fig2}(d)
due to the opening of that channel.

Resonance formation arises from rescattering processes within the
central loop of Fig.~\ref{fig1}, i.e., unitarisation effects.
In this respect, the model has some similarity with the approach of Oset,
Oller, Pel\'{a}ez \em et al.\ \em \cite{DP93}--\cite{NP02}, who take their 
meson-meson Born terms from ChPT, and then unitarise the amplitudes
(see Ref.~\cite{K05a} for an alternative unitarisation scheme).
Note, however, that the present model is unitary from the start, via a
coupled-channel (relativised) Schr\"{o}dinger equation. Moreover, explicit
quark degrees of freedom with a complete confinement spectrum are present,
contrary to ChPT. We also observe that Ref.~\cite{OO99} requires a
``pre-existing'' $s$-channel flavour-singlet resonance at 1021 MeV in order
to reproduce the $f_0(980)$. In our model, such a low-lying $s$-channel
contribution is not needed, only a regular bare $P$-wave $s\bar{s}$ spectrum
from $\approx\!1.5$ GeV upwards. The $f_0(980)$ then appears spontaneously,
though with a $\pi\pi$ width which depends on the bare scalar mixing angle
$\Theta_S$
as well as the dynamical mixing with $\sigma$. The model should also be
contrasted with meson-exchange models (see e.g.\ the work of the J\"{u}lich
group, Janssen \em et al.\ \em \cite{JPHS95}),
in which attraction is generated by a few $t$-channel meson exchanges, whereas
in our approach it arises from the coupling to an infinity of bare $s$-channel
$q\bar{q}$ states. Furthermore, light scalar mesons dynamically
generated in the J\"{u}lich model need not show up as a complete nonet.
Nevertheless, the two models share the feature that the stronger the couplings,
albeit of different natures, the narrower the resonances (see also
below). Most akin to our model is the one of T\"{o}rnqvist \cite{T95}, also
based on bare $s$-channel states coupled to free mesons, which
however does not predict the $\kappa$ meson.

A key point is that phase shifts rise more rapidly as the coupling
constant $\lambda$ increases. Consequently, the masses of the low-lying
resonances move with $\lambda$, but their widths \em decrease \em
\/as $\lambda$ increases.
This is illustrated for the 4 light scalars in Table~\ref{table2}.
\begin{table}[htb]
\begin {center}
\begin{tabular}{r|cccc}
\hline\\[-4mm]
$\lambda$ & $\sigma$ & $\kappa$ & $f_0(980)$ & $a_0(980)$\\\hline\\[-4mm]
 1.5 &  942 - i794 &     ---     &     ---     &     ---      \\ 
 2.0 &  798 - i507 &     ---     &     ---     &     ---      \\ 
 2.2 &  738 - i429 &  791 - i545 &     ---     & 1081 - i8.0  \\ 
 2.4 &  682 - i368 &  778 - i472 &     ---     & 1051 - i25   \\ 
 2.6 &  633 - i319 &  766 - i409 & 1041 - i13  & 1024 - i45   \\ 
 2.8 &  589 - i278 &  754 - i355 & 1028 - i26  &  998 - i61   \\ 
 3.0 &  549 - i243 &  743 - i309 & 1015 - i35  &  978 - i60   \\ 
 3.5 &  468 - i174 &  717 - i219 &  976 - i37  &  896 - i142   \\ 
 4.0 &  404 - i123 &  693 - i155 &  948 - i38  &  802 - i103  \\ 
 5.0 &  308 - i50  &  651 - i69  &  889 - i34  &  711 - i40   \\ 
 7.5 &  216 + i0   &  610 + i0   &  752 - i25  &  632 + i0    \\
10.0 &  142 + i0   &  560 + i0   &  633 - i17  &  577 + i0    \\[-5mm]
\end{tabular}
\end {center}
\caption{Movement of the $\sigma$, $\kappa$, $f_0(980)$ and $a_0(980)$ 
poles as the coupling constant $\lambda$ is varied. Bound states are
indicated by ``+i0''. Units are MeV for the poles and GeV$^{-3/2}$ for
$\lambda$.}
\label{table2}
\end{table}
We see that for very large coupling, the scalars tend to become bound
states (also see Ref.~\cite{BR01}, 3rd paper), while for small coupling
they either disappear to $-i\infty$ in the complex energy plane or become
virtual states.
In this respect, $\sigma$, $\kappa$, $f_0(980)$ and $a_0(980)$ behave
in a completely different fashion to regular $q\bar q$ nonets, (e.g.
$\rho$, $K^*(980)$, etc.) whose resonance widths are directly
proportional to coupling constants.
One of us attempted to fit 10 ratios of coupling constants for $\sigma$,
$\kappa$, $a_0$ and $f_0$ to this conventional scheme \cite{B06_2}.
This attempt failed to reproduce many observed ratios, particularly
for $g^2(\kappa \to K\pi)/g^2(\sigma \to \pi \pi )$ which came out a
factor 3 smaller than experiment.
The reason for the failure is clear: non-perturbative effects due to
the $q\bar q$ loop are crucial.
In the present model, the ratio of $\kappa$ and $\sigma$ widths is
reproduced using closely similar $\lambda$ parameters.

Studying in more detail the trajectory of e.g.\ the $\sigma$ pole,
we see how it emerges from the scattering continuum, passes the
optimised position in Table\ref{table1} for $\lambda\sim3$ GeV$^{-3/2}$,
and then approaches the real axis. However, the pole clearly slows down
for increasing coupling, and very large values of $\lambda$ are needed to
make $\sigma$ a $\pi\pi$ bound state. Here we see the Adler zero in action,
whose influence increases as the pole moves to lower energies. The $\kappa$
pole behaves in a similar way, though it is more difficult to trace for
small $\lambda$ because of the $K\eta'$ Adler zero at $\sim800$ MeV.
Descotes-Genon and Moussallam find a $\kappa$ mass of
$658 \pm 13$~MeV and a width of $\Gamma = 557 \pm 24$~MeV \cite{DGM06}.
Their low mass does not fit well with what we find in Table 2.
We remark that their calculation does not include dynamics of 
the strong coupling of $K\eta '$ to $K\pi$ and consequent mixing
between $K_0^*(1430)$ and $\kappa$.
Below the $K\eta '$ threshold, the analytic continuation of that
channel definitely plays a role, as our calculation verifies.
The $K\eta '$ Adler zero could affect details of their calculation, and
their errors may therefore be too small.

The role of the Adler zero is crucial in understanding $\sigma$,
$\kappa$ and $a_0$.
In particular, the Adler zero for $a_0$ prevents it from having a
mass close to the $\pi\eta$ threshold.
This is why it settles close to the $KK$ threshold, since that channel
has only a distant Adler zero.
We have examined the pole structure of the $\pi\eta$ amplitude.
It is interesting that, as well as having a narrow second sheet
pole due to $a_0(980)$,
there is a very broad pole at $991-i815$ MeV, resembling a bit
the broad poles due to $\sigma$ and $\kappa$;
this broad structure is too wide to determine experimentally.

It is important to test the stability of the fits to data.
Of many such tests, one is to examine the stability of pole
positions over the complete range of parameters shown in Table~\ref{table1}.
This leads to the range of pole positions shown in Table~\ref{table3}.
\begin{table}[htb]
\begin {center}
\begin{tabular}{cc} \hline
Resonance & Range of Pole \\
& Positions (MeV) \\\hline
$\sigma$ & (476--628) - i(226--346) \\
$\kappa $ & (738--902) - i(286--434) \\
$f_0(980)$ & (989--1040) - i(11--38) \\
$a_0(980)$ & (960--1049) - i(23--83) \\ \hline \\[-9mm]
\end{tabular}
\end {center}
\caption{Range of pole positions for the complete range of $\lambda$
and $r_0$ parameters of Table~\ref{table1}; second-sheet poles are
quoted for $f_0$ (only as coupled to $\sigma$) and $a_0$.}
\label{table3}
\end{table}
The poles do not disappear for any parameters within this range.
We have already seen that the $\sigma$ and $\kappa$ poles even survive for
a much wider range of $\lambda$ values alone. So their existence is secure.
However, it is a different story for $f_0(980)$.
If $\lambda$ is reduced to $2.7$ GeV$^{-3/2}$, the cusp at the $KK$
threshold begins to soften, and at $\approx\!2.5$ the pole breaks away from the
$KK$ threshold to become a virtual state at $\approx\!1050$ MeV.

The pole position and line-shape of $a_0$ are likewise sensitive to $\lambda$.
As it is decreased, the cusp at the $KK$ threshold begins to soften for
$\lambda=2.3$ GeV$^{-3/2}$.  The $a_0$ pole breaks away from the $KK$ threshold
and becomes a virtual state at about 1090 MeV for $\lambda\approx2.1$.
It is clear that the $f_0$ and $a_0$ are less tightly bound than $\sigma$ and
$\kappa$. Janssen \em et al.\ \em \cite{JPHS95} have noted that $a_0(980)$
appears as a virtual state in their calculation based on meson exchanges.

The model is undoubtedly an over-simplification; this is deliberate,
so as to expose the essential features.
However, these simplifications lead to difficulties with some
details, which we now discuss.
Firstly, one of the major unknowns afflicting all spectroscopy is
how to continue effects of inelastic channels below their thresholds,
e.g.\ in the processes $KK\to\pi\pi$ and $K\eta'\to K\pi$.
This is relevant to details of both $K\pi$ and $\pi\pi$ phase shifts,
particularly the former.
The model predicts a large coupling of $K_0^*(1430)$ to $K\eta'$, in
agreement with LASS data \cite{A88}.
However, it also predicts a similar large coupling of $\kappa$ to
$K\eta'$; this is because mesons couple via the $q\bar q$ loops of
Fig.~\ref{fig1}.
The combined effect is to predict an even larger amplitude for
$K\eta'\to K\pi$ than is fitted experimentally.
It is important to attentuate this amplitude below the $K\eta'$
threshold, to avoid distorting the fit to the $\kappa$.
The model builds the $K\pi$ Adler zero into the amplitude for
$K\eta' \to K\pi$.
However, the fit improves if we introduce an
additional form factor $\exp (-4|k|^2)$, where $k$ is the magnitude
of the kaon momentum below the inelastic threshold in GeV/c.
This corresponds to the fact that $K_0^*(1430)$ has a long tail of
$K\eta'$ near its threshold. For consistency, the same suppression
factor of subthreshold contributions is applied to the other scalars,
too.

A further technicality is that, coupling only PP channels, too narrow a width
is predicted for $K_0^*(1430)$: $\sim\!200$ MeV, compared with
$294 \pm 23$ MeV quoted by the PDG \cite{PDG04}.
To avoid this conflict, we take the $\kappa$ amplitude from the fit
to experimental data in Ref.~\cite{Ab06}, and fit it only up to 1.2
GeV, where the effect of $K_0^*(1430)$ in our model becomes negligible.
Preliminary results indicate that the inclusion of additional
inelastic channels, such as vector-vector and scalar-scalar, may
improve the widths of higher resonances. This will be the subject of
future work.

Next, the assumption of a sharp radius for the transition
potential is an approximation; this affects radial wave functions.
Thirdly, mesons outside the transition radius are treated as plane
waves, but in reality they will be affected by $t$-channel meson exchanges,
which for simplicity are absent in our approach.

We have fitted $\sigma$ alone and $f_0(980)$ alone, but also
allowed both to couple simultaneously to $\pi\pi$, $KK$, $\eta\eta$,
$\eta\eta'$ and $\eta'\eta'$.
There is obvious mixing between $\sigma$ and $f_0(980)$.
When $\sigma$ and $f_0(980)$ are fitted together, the optimum fit
gives a bare scalar mixing angle of $\sim\!4^\circ$.
However, if $f_0(980)$ is fitted alone with this mixing angle,
the $\pi \pi$ width comes out far too small, $\sim\!2$ MeV, compared
with the observed full width at half maximum of $34 \pm 4$ MeV.
Clearly, the additional dynamical mixing of $f_0(980)$ with $\sigma$, 
mostly through intermediate $KK$, is crucial for a realistic description.

\begin {figure}[htb]
\mbox{} \\[-2.0cm]
\begin {center}
\centerline{\hspace{-2cm}\epsfig{file=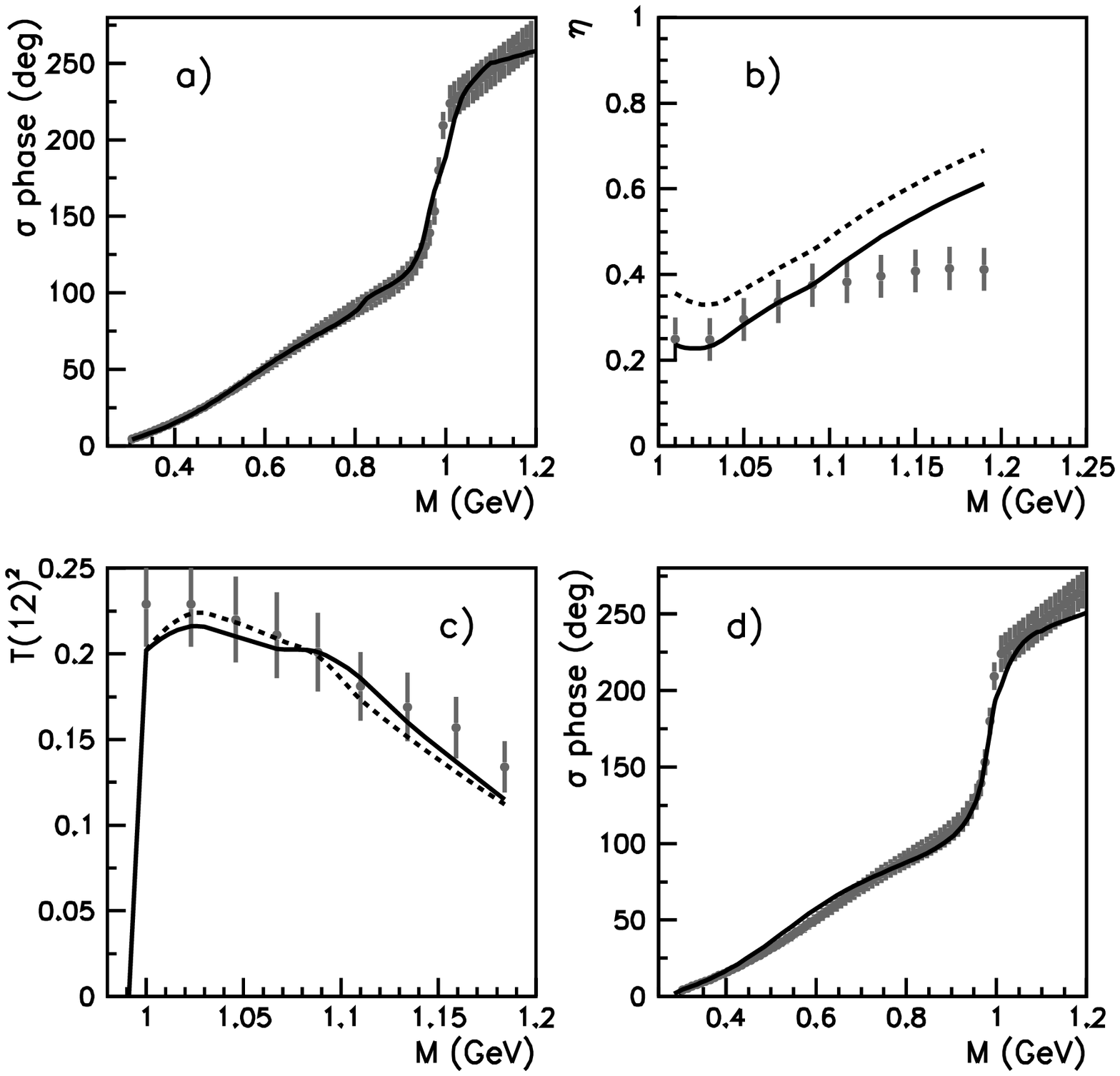,width=15cm}}
\end{center}
\mbox{} \\[-2.5cm]
\caption{Fits to (a) the $\pi \pi$ phase shift with $r_0$,
$\lambda$ and $\theta _S$ fitted to these data alone;
(b) the elasticity parameter $\eta$ for $\pi\pi$ elastic
scattering;
(c) the magnitudes squared of the $T$-matrix for
$\pi \pi \to KK$; and (d) as (a) with $r_0$, $\lambda$ and
$\Theta _S$ fitted to all data simultaneously.}
\label{fig3}
\end {figure}

With the present sharp transition radius, it is difficult to fit
simultaneously:
(i) the elasticity parameter $\eta$ above 1 GeV;
(ii) the intensity of $\pi \pi \to KK$ from 1 to 1.2 GeV;
(iii) $\pi \pi$ phase shifts from 1 to 1.2 GeV.
Any may be fitted alone, but the threefold combination exposes
some conflicts.
Figure~\ref{fig3} shows two fits.
In (a), we show the recent $\pi \pi$ phases of Kaminski,
Pel\'{a}ez and Yndurain~\cite{KPY06} above 1 GeV.
Figure~\ref{fig3}(a) and full lines on (b) and (c) show fits to individual
sets of data, varying $\lambda$, $r_0$ and $\theta _S$.
This achieves a good fit to all sets of data.
Figure~\ref{fig3}(d) and the dashed lines on Figs.~\ref{fig3}(b) and (c)
show the best fit when all three sets of data are fitted simultaneously.
The pole position of $f_0(980)$ is then at $1007 -i38$ MeV, compared
with the experimental value $(998 \pm 4) - i(18 \pm 4)$ from BES
parameters.

The optimum value of $\Theta_S$ is $\sim\!4^\circ$, quite a bit smaller than
frequently quoted values (see e.g.\ Ref.~\cite{DS98}). However, fitting
$f_0(980)$ alone gives rise to a much larger $\Theta_S$, in excess of
$30^\circ$. So there is a sizable, $s$-dependent dynamical
contribution to the scalar mixing angle, which is largest in the vicinity of
the $f_0(980)$, that is, near the $KK$ threshold. This is a physically
appealing result of our fits in this sector.

A further detail concerns the pseudoscalar mixing angle
$\Theta_{PS}$.
The experimental value from Crystal Barrel is $(-17.3 \pm 1.8)^\circ$
\cite{Am92}; that is the value used here (also see Ref.~\cite{K05b}).
However, the fit to the line-shape of $a_0(980) \to \eta \pi$
is sensitive to $\Theta_{PS}$.
For those data, the optimum value is $\Theta_{PS} = -(7 \pm 3)^\circ$.
This value is used for Fig.~\ref{fig2}(b).
With $\Theta_{PS} = -17.3^\circ$, the fit is 15\% too high on both
sides of the peak, as shown by the dotted curve.
These discrepancies might be associated with the extensive $KK$ cloud
around $a_0(980)$ or with the $qq\bar q\bar q$ correlations proposed by Jaffe.
Alternatively, small yet non-negligible contributions from higher, closed
thresholds may improve the predicted $\Theta_{PS}$, as indicated by
preliminary fits. 

Despite the blemishes described above, the essential results are as
follows. The model of Fig.~\ref{fig1} is capable of generating all of
$\sigma$, $\kappa$, $a_0(980)$ and $f_0(980)$ dynamically
via the intermediate $q\bar q$ loop. This is a non-perturbative effect.
The overall coupling constant required for the four resonances varies
by only $\pm 9\%$ and the radius parameter $r_0$ by $\pm10\%$. Moreover,
all scalar poles survive, with reasonable values, over the complete range
of fitted parameters, so the model is very robust.
The widths of these dynamical resonances decrease as the coupling constant
increases. This is an essential difference from regular $q\bar{q}$ states,
where, for not too large coupling, widths are roughly \em proportional \em
\/to the square of the coupling.
Nevertheless, the formation of these four resonances cannot been
dissociated from the presence of a bare $q\bar{q}$ spectrum, which
is crucial for the generation of dynamical resonances with moderate
widths. In particular, for each ground-state bare scalar $q\bar{q}$ state,
a pair of resonances is produced by coupling to the meson-meson continuum,
namely one light, non-standard scalar meson, and one regular scalar in
the energy region 1.3--1.5 GeV. This phenomenon was first observed two
decades ago (see 2nd paper of Ref.~\cite{BRMDRR86}), and later confirmed,
to some extent, by T\"{o}rnqvist \cite{T95}.

\section*{Acknowledgments}

This work was supported by the {\it Funda\c{c}\~{a}o para a Ci\^{e}ncia
e a Tecnologia} \/of the {\it Minist\'{e}rio da Ci\^{e}ncia, Tecnologia
e Ensino Superior} \/of Portugal, under contracts POCTI/FP/FNU/50328/2003
and POCI/FP/63437/2005. One of us (FK) also acknowledges partial support
from grant SFRH/ BPD/9480/2002  and the Czech project LC06002.


\begin{thebibliography} {99}
\bibitem{Ai01} 
E.~M.~Aitala {\it et al.}  [E791 Collaboration],
Phys.\ Rev.\ Lett.\  {\bf 86} (2001) 765
[arXiv:hep-ex/0007027].

\bibitem{Ab04} 
M.~Ablikim {\it et al.}  [BES Collaboration],
Phys.\ Lett.\ B {\bf 598} (2004) 149
[arXiv:hep-ex/0406038].

\bibitem{Ai02} 
E.~M.~Aitala {\it et al.}  [E791 Collaboration],
Phys.\ Rev.\ Lett.\  {\bf 89} (2002) 121801
[arXiv:hep-ex/0204018].

\bibitem{Ai06}
E.~M.~Aitala {\it et al.}  [E791 Collaboration],
Phys.\ Rev.\ D {\bf 73} (2006) 032004
[arXiv:hep-ex/0507099].

\bibitem{B05} 
D.~V.~Bugg,
Eur.\ Phys.\ J.\ A {\bf 25} (2005) 107
[Erratum-ibid.\ A {\bf 26} (2005) 151]
[arXiv:hep-ex/0510026].

\bibitem{Ab06} 
M.~Ablikim {\it et al.}  [BES Collaboration],
Phys.\ Lett.\ B {\bf 633} (2006) 681
[arXiv:hep-ex/0506055].

\bibitem{B06_1} 
D.~V.~Bugg,
Phys.\ Lett.\ B {\bf 632} (2006) 471
[arXiv:hep-ex/0510019].

\bibitem{Ab05} 
M.~Ablikim {\it et al.}  [BES Collaboration],
Phys.\ Lett.\ B {\bf 607} (2005) 243
[arXiv:hep-ex/0411001].

\bibitem{J77} 
R.~L.~Jaffe,
Phys.\ Rev.\ D {\bf 15} (1977) 267.

\bibitem{S82}
M.~D.~Scadron,
Phys.\ Rev.\ D {\bf 26} (1982) 239.

\bibitem{BRMDRR86} 
E.~van Beveren, G.~Rupp, T.~A.~Rijken and C.~Dullemond,
Phys.\ Rev.\ D {\bf 27}, 1527 (1983);
E.~van Beveren, T.~A.~Rijken, K.~Metzger, C.~Dullemond, G.~Rupp and J.~E.~Ribeiro,
Z.\ Phys.\ C {\bf 30} (1986) 615;
E.~van Beveren and G.~Rupp,
Phys.\ Rev.\ Lett.\  {\bf 93} (2004) 202001
[arXiv:hep-ph/0407281].

\bibitem{BR01} 
E.~van Beveren and G.~Rupp,
Eur.\ Phys.\ J.\ C {\bf 22} (2001) 493
[arXiv:hep-ex/0106077];
E.~van Beveren and G.~Rupp,
Int.\ J.\ Theor.\ Phys.\ Group Theor.\ Nonlin.\ Opt.\ {\bf11} (2006) 179
[arXiv:hep-ph/0304105];
E.~van Beveren and G.~Rupp,
Phys.\ Rev.\ Lett.\  {\bf 91} (2003) 012003
[arXiv:hep-ph/0305035];
E.~van Beveren, F.~Kleefeld and G.~Rupp,
AIP Conf.\ Proc.\  {\bf 814} (2006) 143
[arXiv:hep-ph/0510120].

\bibitem{BR04}
E.~van Beveren and G.~Rupp,
Mod.\ Phys.\ Lett.\ A {\bf 19} (2004) 1949
[arXiv:hep-ph/0406242];
E.~van Beveren, J.~E.~G.~Costa, F.~Kleefeld and G.~Rupp,
Phys.\ Rev.\ D {\bf 74} (2006) 037501
[arXiv:hep-ph/0509351].

\bibitem{AKPSS00} 
V.~V.~Anisovich, A.~A.~Kondashov, Y.~D.~Prokoshkin, S.~A.~Sadovsky and A.~V.~Sarantsev,
Phys.\ Atom.\ Nucl.\  {\bf 63} (2000) 1410
[Yad.\ Fiz.\  {\bf 63} (2000) 1410]
[arXiv:hep-ph/9711319].

\bibitem{RKB05}
G.~Rupp, F.~Kleefeld and E.~van Beveren,
AIP Conf.\ Proc.\  {\bf 756} (2005) 360
[arXiv:hep-ph/0412078];
F.~Kleefeld,
AIP Conf.\ Proc.\  {\bf 717} (2004) 332
[arXiv:hep-ph/0310320].

\bibitem{B03}
D.~V.~Bugg,
Phys.\ Lett.\ B {\bf 572} (2003) 1
[Erratum-ibid.\ B {\bf 595} (2004) 556].

\bibitem{H73} 
B.~Hyams {\it et al.},
Nucl.\ Phys.\ B {\bf 64} (1973) 134
[AIP Conf.\ Proc.\  {\bf 13} (1973) 206].

\bibitem{P01} 
S.~Pislak {\it et al.}  [BNL-E865 Collaboration],
Phys.\ Rev.\ Lett.\  {\bf 87} (2001) 221801
[arXiv:hep-ex/0106071].

\bibitem{CGL01} 
G.~Colangelo, J.~Gasser and H.~Leutwyler,
Nucl.\ Phys.\ B {\bf 603} (2001) 125
[arXiv:hep-ph/0103088].

\bibitem{A88} 
D.~Aston {\it et al.},
Nucl.\ Phys.\ B {\bf 296} (1988) 493.

\bibitem{B06_2} 
D.~V.~Bugg,
Eur.\ Phys.\ J.\ C {\bf 47} (2006) 45
[arXiv:hep-ex/0603023].

\bibitem{Al02} 
A.~Aloisio {\it et al.}  [KLOE Collaboration],
Phys.\ Lett.\ B {\bf 536} (2002) 209
[arXiv:hep-ex/0204012].

\bibitem{BASZ94} 
D.~V.~Bugg, V.~V.~Anisovich, A.~Sarantsev and B.~S.~Zou,
Phys.\ Rev.\ D {\bf 50} (1994) 4412.

\bibitem{K05b}
F.~Kleefeld,
Acta Physica Slovaca {\bf 56} (2006) 373
[arXiv:nucl-th/0510017].

\bibitem{DP93} 
A.~Dobado and J.~R.~Pel\'{a}ez,
Phys.\ Rev.\ D {\bf 47} (1993) 4883
[arXiv:hep-ph/9301276].

\bibitem{OO97} 
J.~A.~Oller and E.~Oset,
Nucl.\ Phys.\ A {\bf 620} (1997) 438
[Erratum-ibid.\ A {\bf 652} (1999) 407]
[arXiv:hep-ph/9702314].

\bibitem{OOP99} 
 J.~A.~Oller, E.~Oset and J.~R.~Pel\'{a}ez,
Phys.\ Rev.\ D {\bf 59} (1999) 074001
[Erratum-ibid.\ D {\bf 60} (1999) 099906]
[arXiv:hep-ph/9804209].

\bibitem{OO99} 
J.~A.~Oller and E.~Oset,
Phys.\ Rev.\ D {\bf 60} (1999) 074023
[arXiv:hep-ph/9809337].

\bibitem{JOP00} 
M.~Jamin, J.~A.~Oller and A.~Pich,
Nucl.\ Phys.\ B {\bf 587} (2000) 331
[arXiv:hep-ph/0006045].

\bibitem{NP02} 
A.~Gomez Nicola and J.~R.~Pel\'{a}ez,
Phys.\ Rev.\ D {\bf 65} (2002) 054009
[arXiv:hep-ph/0109056].

\bibitem{K05a}
F.~Kleefeld,
PoS {\bf HEP2005} (2006) 108
[arXiv:hep-ph/0511096].

\bibitem{JPHS95} 
G.~Janssen, B.~C.~Pearce, K.~Holinde and J.~Speth,
Phys.\ Rev.\ D {\bf 52} (1995) 2690
[arXiv:nucl-th/9411021]. \\
Also see
V.~Baru, J.~Haidenbauer, C.~Hanhart, Y.~Kalashnikova and A.~Kudryavtsev,
Phys.\ Lett.\ B {\bf 586} (2004) 53
[arXiv:hep-ph/0308129].

\bibitem{T95}
N.~A.~T\"{o}rnqvist,
Z.\ Phys.\ C {\bf 68}, 647 (1995)
[arXiv:hep-ph/9504372];
N.~A.~T\"{o}rnqvist and M.~Roos,
Phys.\ Rev.\ Lett.\  {\bf 76}, 1575 (1996)
[arXiv:hep-ph/9511210].

\bibitem{DGM06}
S.~Descotes-Genon and B.~Moussallam,
arXiv:hep-ph/0607133.

\bibitem{PDG04} 
S.~Eidelman {\it et al.}  [Particle Data Group],
Phys.\ Lett.\ B {\bf 592} (2004) 1.

\bibitem{KPY06} 
R.~Kaminski, J.~R.~Pel\'{a}ez and F.~J.~Yndurain,
arXiv:hep-ph/0603170.

\bibitem{DS98}
R.~Delbourgo and M.~D.~Scadron,
Int.\ J.\ Mod.\ Phys.\ A {\bf 13} (1998) 657
[arXiv:hep-ph/9807504].

\bibitem{Am92} 
C.~Amsler {\it et al.}  [Crystal Barrel Collaboration],
Phys.\ Lett.\ B {\bf 294} (1992) 451.

\end{thebibliography}
\end {document}